\documentclass[%
 reprint,
 superscriptaddress,
 showkeys,
 amsmath,amssymb,
 aps,
]{revtex4-2}

\usepackage[utf8]{inputenc}
\usepackage[T1]{fontenc}
\usepackage[german,english]{babel}
\usepackage{csquotes}
\usepackage[pdftex]{graphicx}
\usepackage{xspace}
\usepackage{multirow}
\usepackage{dcolumn}
\usepackage{bm}
\newcommand{\EQE}{\mathrm{EQE_{PV}}}
\newcommand{\EL}{\mathrm{EQE_{EL}}}
\newcommand{\reorg}{\lambda_\mathrm{R}}
\newcommand{\disord}{\sigma_{\mathrm{ct}}}
\newcommand{\kT}{k_\mathrm{B}T}
\newcommand{\Ect}{E_\mathrm{ct}}
\newcommand{\Voc}{V_\mathrm{oc}}
\newcommand{\FC}{\frac{e^{-S} S^j}{j!}}
\newcommand{\hw}{ \Lambda_\mathrm{vibr} } 
\newcommand{\hwLow}{ \Lambda_\mathrm{2} } 
\newcommand{\f}{f^{*}}
\newcommand{\fulleren}{C$_{60}$}
\newcommand{\Tvalid}{T_\mathrm{valid}}
\newcommand{\Tset}{T_\mathrm{set}}
\newcommand{\LED}{\mathrm{EQE_{LED}}}

\newcommand{\tapc}{TAPC$_\mathrm{10\%}$:\fulleren}
\newcommand{\simpleMarcus}{\emph{simple Marcus model}\xspace}
\newcommand{\extdMarcus}{\emph{extended Marcus model}\xspace}
\newcommand{\disorderedMarcus}{\emph{disordered Marcus model}\xspace}
\newcommand{\extdDisorderedMarcus}{\emph{extended disordered model}\xspace}
\newcommand{\multiVibrations}{\emph{multiple vibrations model}\xspace}
\newcommand{\IQE}{\mathrm{IQE_{PV}}}
\newcommand{\unit}[1]{\mathrm{\, #1}}
\begin{document}

%\preprint{APS/123-QED}
\title{Temperature dependent charge transfer state absorption and emission reveal dominant role of dynamic disorder in organic solar cells}

\author{Clemens Göhler}
\author{Maria Saladina}
\affiliation{%
	Institut für Physik, Technische Universität Chemnitz, 09126 Chemnitz, Germany
}%

\author{Yazhong Wang}
\author{Donato Spoltore}
\author{Johannes Benduhn}
\author{Karl Leo}
\affiliation{
	Dresden Integrated Center for Applied Physics and Photonic Materials (IAPP) and Institute for Applied Physics, Technische Universität Dresden, Nöthnitzer Str. 61, 01187 Dresden, Germany
}%

\author{Carsten Deibel}
\email{deibel@physik.tu-chemnitz.de}
\affiliation{%
	Institut für Physik, Technische Universität Chemnitz, 09126 Chemnitz, Germany
}%

\date{\today}

\begin{abstract}
The energetic landscape of charge transfer (CT) states at the interface of electron donating and electron accepting domains in organic optoelectronic devices is crucial for their performance. Central questions---such as the role of static energetic disorder and vibrational effects---are under ongoing dispute. This study provides an in-depth analysis of temperature dependent broadening of the spectroscopic absorption and emission features of CT states in devices with small molecule--fullerene blends. We confirm the validity of the electro-optical reciprocity relation between the photovoltaic external quantum efficiency ($\EQE$) and electroluminescence ($\EL$), enabling us to validate the device temperature during the experiment. The validated temperature allows us to fit our experimental data with several models, and compare extracted CT state energies with the corresponding open circuit voltage limit at $0\unit{K}$. Our findings unveil that the absorption and emission characteristics are usually not symmetric, and dominated by temperature-activated broadening (vibrational) effects instead of static disorder. 
\end{abstract}

\maketitle

\section{Introduction}\label{sec:introduction}

Organic solar cells (OSCs) gain renewed interest since their power conversion efficiency increased rapidly over the last few years and are nowadays approaching 20\%.\cite{ma2020achieving, cui2020single, liu202018} Many successful steps have been achieved in designing new materials, especially non-fullerene acceptors; yet the underlying working principles remain relatively unchanged.\cite{gohler2018nongeminate, gurney2019review, an2020recent, ma2020recent} All efficient approaches favor a combination of at least two materials, namely an electron donor and acceptor, forming a heterogeneously blended layer sandwiched between selective electrodes. To facilitate the dissociation of strongly bound photogenerated excitons usually free energy is sacrificed. This is realized at the donor--acceptor interface by the formation of energetically more favorable and less strongly bound charge transfer (CT) states. 

On the one hand, CT states are responsible for more efficient charge carrier generation. On the other hand, they are limiting the open circuit voltage $\Voc$ in organic solar cells\cite{vandewal2008relation} and determining both radiative and non-radiative loss mechanisms.\cite{vandewal2010relating, burke2015beyond} Nevertheless, a detailed description of their energetic nature is still missing. The majority of studies in the field over the last few years determined the energetic properties utilizing (electro-)optical spectroscopy and interpreted the results based on a molecular model of CT states,\cite{vandewal2010relating, vandewal2008relation, kahle2018interpret, azzouzi2018nonradiative, melianas2019nonequilibrium, collado2019energy, burke2015beyond, benduhn2017intrinsic, vandewal2017charge, tvingstedt2020temperature} and recently computational multi-scale simulations together with quantum-mechanical calculations.\cite{panhans2020molecular} One of the most discussed questions is whether CT states are dominated by static energetic disorder at the donor--acceptor interface, which we would expect in organic semiconductor blends based on conclusions from trap state\cite{schafferhans2010oxygen} and charge transport measurements,\cite{blakesley2011relationship,street2011photoconductivity} or can be described by dynamic vibrational broadening alone. As this matter impacts the interpretation of measured device properties, including the excited CT state energy $\Ect$, it is essential to precisely quantify energy losses in organic solar cells.\cite{burke2015beyond, vandewal2008relation, vandewal2009origin, kahle2018interpret} A related question is whether $\Voc$ losses measured in a specific device are inherent to a material system, or can be potentially reduced by optimizing the morphology in order to decrease static disorder.

In this work, we address the role of energetic disorder by investigating the validity of different theoretical approaches. We compare the predicted absorption and emission spectra to our experimental photovoltaic external quantum efficiency ($\EQE$) and electroluminescence ($\EL$) spectra, measured on a set of thermally evaporated OSCs over a broad temperature range. Our measurements provide evidence for the validity of the electro-optical reciprocity between $\EQE$ and $\EL$, being essential to confirm (or correct) the device temperature during experiments. We find that the \emph{multiple vibrations model}, which does not include static disorder, explains our experimental results the best. In order to optimize $\Voc$ in the absence of static disorder, the emphasis should shift to the reduction of vibration-induced losses.\cite{panhans2020molecular} 

\section{Background Information}

\subsection{Electro-optical reciprocity relation}

As optical absorption of CT states is usually very weak in OSCs, it is often measured by sensitive detected $\EQE$ spectroscopy. Absorption and $\EQE$ are related by the internal photovoltaic quantum efficiency ($\IQE$) for conversion of absorbed photons to extracted charge carriers. Emission from CT states is quantified by spectral electroluminescence (EL) yield ($\EL$) under forward bias. Its integral value $\LED$, the total EL quantum efficiency, represents all non-radiative voltage losses of the photovoltaic device, which scale with $\ln(\LED)$.\cite{vandewal2010relating} While CT state emission can also be observed in terms of photoluminescence,\cite{kahle2018interpret} it may be more difficult due to much stronger emission from pure donor and acceptor phases which have to be subtracted first.\cite{tvingstedt2020temperature} Optical interference might also affect the light out-coupling efficiency of thin film devices, which was recently pointed out by List~\textit{et al.}\cite{list2018correct} The effect was shown to be negligible for absorbing layer thicknesses below $80\unit{nm}$,\cite{list2018correct, melianas2019nonequilibrium, armin2020limitations} and to be independent of the device temperature.\cite{tvingstedt2020temperature}

$\EQE$ and $\EL$ of semiconductor devices are connected by Rau's electro-optical reciprocity relation.\cite{rau2007reciprocity} Assuming an $\IQE$ independent of the energy of the absorbed photon\cite{vandewal2014efficient} and negligible space-charge effects,\cite{kirchartz2016reciprocity} the emission flux from a solar cell operated under forward bias is proportional to its thermal emission spectrum: both spectral distributions ($\EQE$ \& $\EL$) are correlated by the blackbody emission flux $\phi_\mathrm{BB}(E,T)$ of photons with energy $E$ at temperature $T$, and scale with the dark saturation current $J_0$ of the device. 
The validity of the reciprocity relation was shown to hold for most types of photovoltaic devices, including OSCs at room temperature.\cite{vandewal2010relating, yao2015quantifying, muller2014effect, melianas2019nonequilibrium} Therefore, a theoretical description of CT state emission and absorption characteristics is required to satisfy Rau's reciprocity relation: 

\begin{eqnarray}
    \EL(E,T) ~\mathrm{d}E &&= \EQE(E,T)\phi_\mathrm{BB}(E,T) \frac{q~ \mathrm{d}E}{J_0(T)} \label{eq:rauReciprocity} 
\end{eqnarray}

where $q$ is the elementary charge. The total EL efficiency $\LED(T)$ is given by the integral of either side of equation~(\ref{eq:rauReciprocity}). 

In the experiment, the temperature of the device under test is usually set to the temperature of a reservoir $\Tset$, for example by using a PID controlled heater in an insulated cryostat. Most studies reporting temperature-dependent investigations discuss their findings in terms of $\Tset$,\cite{tvingstedt2020temperature, linderl2020crystalline, burke2015beyond} without verifying if that temperature is actually valid under the measurement conditions (e.g.\ current injected to drive the electroluminescence, which can heat up the sample). If the reciprocity relation holds over a broad temperature range, it allows us to validate the solar cell temperature $\Tvalid$ during spectroscopic measurements: by using the Boltzmann approximation of the blackbody spectrum, the dark saturation current $J_0$ and $\Tvalid$ can be extracted from the reciprocity relation and the quotient of $\EL$ and $\EQE$.\cite{vandewal2014efficient} With the Boltzmann constant $k_\mathrm{B}$, the speed of light in vacuum $c$, and Planck's constant $h$, we find:

\begin{eqnarray}
    \ln{\left(\frac{q}{J_0}\right)} &&- \frac{E}{k_\mathrm{B}\Tvalid} = \label{eq:inverseReciprocity}\\
    &&-2 \ln{\left(E\right)} + \ln{\left(\frac{c^2h^3}{2\pi}\right)} + \left.\ln{\left(\frac{\EL(E)}{\EQE(E)}\right)}\right|_{\Tset} \nonumber
\end{eqnarray}

Employing equation~(\ref{eq:inverseReciprocity}), it can be tested if the device really is at $\Tset$ during the absorption and during the emission measurement. We calculate $\Tvalid$ of the gray body that would fulfill the reciprocity relation of measured $\EL(\Tset)$ and $\EQE(\Tset)$. If we find $\Tvalid \neq \Tset$, either one or both spectra were recorded at a device temperature other than $\Tset$. The crucial parameter in these cases is most likely the device temperature during $\EL$ measurements, as injection conditions exceeding typical working parameters may cause unintended heating of the device. Monochromatic light intensities for $\EQE$ spectroscopy are typically far below solar illumination and should not cause unintended heating; an approximation of the power dissipated by the devices during $\EL$ and $\EQE$ measurements is given in sec.~2 in the Supplemental material. Accordingly, high injection currents have been shown to affect shapes of $\EL$ spectra,\cite{linderl2020crystalline} and to cause deviations from the electro-optical reciprocity relation even at room temperature.\cite{vandewal2010relating, vandewal2014efficient} As a consequence of equation~(\ref{eq:inverseReciprocity}), we expect that the device temperature for $\EQE$ measurements is usually very close to $\Tset$, and the temperature during $\EL$ measurements is often $\Tvalid \ge \Tset$ due to Joule heating caused by the higher current densities. We therefore suggest to estimate the device temperature with $\Tvalid$ for the analysis of emission characteristics.

\subsection{CT state models}\label{sec:models}

Currently discussed theoretical descriptions of CT states are based on the theory of electron transfer by Marcus.\cite{marcus1989relation} The interfacial CT state is treated in analogy to a molecular state, with photon absorption and emission from excited electronic states. An application of this theory with regard to the electro-optical properties of a solar cell device was given by Vandewal \textit{et al.},\cite{vandewal2010relating,vandewal2008relation} who were able to explain both the sub-bandgap characteristics of $\EQE$ and $\EL$ of OSCs at room temperature with Gaussian line shapes, 

\begin{eqnarray}
    \EQE = \frac{ E^{-1}\times f_\sigma^*}{\sqrt{4\pi \reorg \kT}} &&\exp{\left( - \frac{(\Ect + \reorg - E)^2}{4 \reorg \kT } \right)} \label{eq:vdw_EQE} \\
    \EL = \frac{ E \times \f_\mathrm{emis} / \phi_\mathrm{inj} }{\sqrt{4\pi \reorg \kT}} &&\exp{\left( - \frac{(\Ect - \reorg - E)^2}{4 \reorg \kT} \right) } .\label{eq:vdw_EL} 
\end{eqnarray}

The absorption and emission lines of the CT state with energy $\Ect$ are separated by twice its reorganization energy $\reorg$, which also describes the respective linewidths $(2\reorg\kT)^{1/2}$; only these two parameters are necessary to describe the shape of the contribution of CT states to $\EL$ and $\EQE$ spectra at temperature $T$. Further assumptions and quantum mechanical photo-physics, including transition matrix elements, are merged into the amplitudes $f_\mathrm{\sigma}^{*}$ and $f_\mathrm{emis}^{*}$. The emission flux under LED conditions has to be normalized by the injected charge carrier flux $\phi_\mathrm{inj}$. Using this model, the authors were able to predict $\Voc$ of the solar cell under simulated solar illumination at room temperature from the extracted $\Ect$.\cite{vandewal2010relating, burke2015beyond} 

In this \simpleMarcus, $\reorg$ accounts for dynamic broadening from intra- and intermolecular vibrations with small characteristic frequencies $f_\mathrm{vibr}\ll\kT/h$. Higher frequency vibrations inherent to organic molecules such as carbon--carbon stretching modes, which have been proposed to play an important role for non-radiative recombination in organic semiconductor devices,\cite{benduhn2017intrinsic} lead to additional absorption and emission lines.\cite{gould1993radiative, kahle2018interpret} For a single dominant vibration mode, the corresponding spectra can be explained by a sum of shifted Gaussian lines, leading to an \extdMarcus:

\begin{eqnarray}
    \EQE~ && \propto \frac{E^{-1}}{\sqrt{4\pi \reorg \kT}} \nonumber\\
    \times\sum_{j=0}^\infty &&\FC \exp\left(-\frac{(\Ect + j\hw + \reorg - E)^2}{4\reorg\kT}\right) \label{eq:gould_EQE}
\end{eqnarray} 

The moment of the Poisson distribution characterized by the Huang--Rhys factor $S$ determines the relative intensity of vibrational levels $j$. For vibration energies $\hw>100\unit{meV}$, these contributions may not be visible in the $\EQE$ spectrum due to the overlying stronger absorption from donor and/or acceptor singlet states.\cite{kahle2018interpret} Characteristic carbon--carbon stretching mode energies of fullerenes were found in the region of $\hw\approx150\ldots180\unit{meV}$.\cite{tvingstedt2020temperature,kahle2018interpret, fuchs2017vibrational} Note that excitation of phonons leads to additional absorption at higher energies, whereas the respective photoluminescence spectrum features additional emission lines redshifted by $-j\hw$.\cite{kahle2018interpret, gould1993radiative}

Equation~(\ref{eq:gould_EQE}) originates from the extended Marcus--Levich--Jortner model of the electron transfer theory. In this context, $\hw$ represents one discrete intramolecular \textit{quantum mode} of the charge transfer complex, and $\reorg$ is associated with various optical phonons of the surrounding medium, which can be treated as classical due to their mean vibration energy being much smaller than $\kT$.\cite{jortner1976temperature, ulstrup1975effect}

Without taking these high-energy modes into account, the \simpleMarcus~(equations~(\ref{eq:vdw_EQE}) and~(\ref{eq:vdw_EL})) complies with the reciprocity relation, equation~(\ref{eq:rauReciprocity}). As a consequence, the reduced spectra ($\EQE\cdot E$ and $\EL/E$) are symmetric and mirror each other. However, when the electro-optical reciprocity relation is applied to the \textit{extended Marcus model's} $\EQE$ (equation~(\ref{eq:gould_EQE})), we can derive a different expression for the $\EL$ emission:

\begin{eqnarray}
    &&\EL \propto \frac{E}{\sqrt{4\pi\reorg\kT}} \exp{\left(-\frac{\Ect}{\kT}\right)} \times \label{eq:gould_EL}\\
    && \sum_{j=0}^\infty \FC ~e^{\left(-j\frac{\hw}{\kT}\right)} \exp\left(-\frac{(\Ect + j\hw - \reorg-E)^2}{4\reorg\kT}\right) . \nonumber
\end{eqnarray}

We find that the $\EL$ does not longer mirror $\EQE$, as it inherits the positive vibrational progression $+j\hw$. The emission quantum efficiency from these states is however exponentially damped by the term $\exp\left(-j \hw / \kT\right)$, yielding an asymmetrical shape of the reduced spectra, in contrast to the \simpleMarcus. 

\subsection{Models based on static and dynamic disorder}

Both models discussed above do not include static energetic disorder, which is expected to be present in non-crystalline organic semiconductors, and even more so in those made from heterogeneously blended films. An extension to the $\EQE$ description from the \simpleMarcus\ was given by Burke \textit{et al.}\ by assuming a Gaussian distribution of disordered excited CT state energies $\Ect'$ with variance $\disord^2$ to represent static disorder. With this assumption, the $\EQE$ shape can still be described by a reduced Gaussian, albeit with a modified variance and $\Ect$ now representing the first moment of the distribution of CT state energies:\cite{burke2015beyond}

\begin{eqnarray}
    \EQE \propto &&\frac{E^{-1}\sqrt{\pi^{-1}}}{\sqrt{4\reorg\kT+2\disord^2}}~ \exp\left(-\frac{(\Ect + \reorg - E)^2}{4\reorg\kT + 2\disord^2}\right)\nonumber \\ ~ \label{eq:burke_EQE} \\ 
    \EL \propto &&\frac{E\sqrt{\pi^{-1}}}{\sqrt{4\reorg\kT+2\disord^2}}  \exp{\left(-\frac{\Ect}{\kT}+\frac{\disord^2}{2(\kT)^2}\right)}\nonumber\\
    &&~~\times\exp\left(-\frac{(\Ect-\reorg-\disord^2/\kT-E)^2}{4\reorg\kT + 2\disord^2}\right) .  \label{eq:burke_EL}  
\end{eqnarray} 

The EL expression in this \disorderedMarcus\ given here complies with $\EQE$~(equation~(\ref{eq:burke_EQE})) and the electro-optical reciprocity relation (equation~(\ref{eq:rauReciprocity})). In both the $\EQE$ and $\EL$ spectra, the modified variance $\sigma^2=\disord^2 + 2\reorg\kT$ leads to a non-zero linewidth at $T=0\unit{K}$, a trend that was experimentally confirmed by several temperature-dependent measurements of CT states in $\EQE$ or $\EL$ spectra.\cite{tvingstedt2020temperature, burke2015beyond,linderl2020crystalline,panhans2020molecular}

The model predicts an additional temperature dependent peak redshift of the $\EL$ spectrum by $-\disord^2/\kT$, while reduced spectra should nonetheless remain symmetric with equal $\EQE$ and $\EL$ linewidths. This particular peak shift should only be affected by the device temperature; it has been reported for inorganic solar cells with a Gaussian distribution of bandgaps,\cite{kirchartz2007electroluminescence} but is absent in many studies on OSCs.\cite{tvingstedt2020temperature, linderl2020crystalline} Nevertheless, prominent EL peak blueshifts caused by increased injection currents have been documented for OSCs with an exponential\cite{linderl2020crystalline} and, less pronounced, with a Gaussian distribution of sub-bandgap states.\cite{gong2012influence, melianas2019nonequilibrium, linderl2020crystalline} 

\begin{table}[tb]
\footnotesize
\setlength{\tabcolsep}{10pt} 
\renewcommand{\arraystretch}{1.5} 
\caption{\label{tab:table1}Comparison of the models included in this study, separated for inclusion of static disorder and phonon excitation. The expected symmetry between reduced absorption and emission lines is included as an additional criterion. Allowing excitation of molecular vibrations disrupts the symmetry except for the 0--0 transition; incorporating static energetic disorder introduces a temperature dependent shift of the $\EL$ peak.}
\begin{ruledtabular}
\setlength{\tabcolsep}{0em}
\begin{tabular}{lll}
 & w/o molecular vibrations & w/ molecular vibrations \\ \hline
 \multirow{9}{*}{\rotatebox[origin=c]{90}{\parbox[c]{4cm}{\centering w/o static disorder}}} & \textbf{simple Marcus} & \textbf{extended Marcus} \\
 & \textbf{model~\footnote[1]{symmetrical reduced $\EQE$ \& $\EL$ spectra}} ~[Vandewal \textit{et al.}] & \textbf{model~\footnote[2]{symmetrical only on 0-0 transition line of reduced spectra}\footnote[3]{asymmetrical envelope function}} \\ \cline{2-3}
 & $\Ect = E_\mathrm{PV,max} - \reorg$ & $\Ect = E_\mathrm{PV,max}^{~0-0} - \reorg$\\
 & $\Ect = E_\mathrm{EL,max} + \reorg$ & $\Ect = E_\mathrm{EL,max}^{~0-0} + \reorg$ \\
 & $\sigma_\mathrm{PV,EL}=\sqrt{2\reorg\kT}$ & $\sigma_\mathrm{PV,EL}^{0-0}=\sqrt{2\reorg\kT}$ \\ \cline{2-3} 
 & & \textbf{multiple vibrations~\footnote[4]{asymmetrical reduced $\EQE$ \& $\EL$ spectra, depending on vibrations involved}} \\ \cline{3-3}
 & & $E_\mathrm{PV,max} = f(\Ect, \reorg) $\\
 & & $E_\mathrm{EL,max} = f(\Ect, \reorg, T)$ \\
 & & $\sigma_\mathrm{PV} \neq \sigma_\mathrm{EL}$ \\ \hline 
 \multirow{5}{*}{\rotatebox[origin=c]{90}{\parbox[c]{2.68cm}{\centering w/ static disorder}}} & \textbf{disordered Marcus} & \textbf{extended disordered} \\
 & \textbf{model~\footnotemark[1]\footnote[5]{symmetry axis redshifted with decreasing temperature}}~[Burke \textit{et al.}] & \textbf{model~\footnotemark[2]\footnotemark[5]}~[Kahle \textit{et al.}] \\ \cline{2-3}
  & $\Ect = E_\mathrm{PV,max} - \reorg$ & $\Ect = E_\mathrm{PV,max}^{~0-0} - \reorg$\\
 & $\Ect = E_\mathrm{EL,max} + \reorg + \frac{\disord^2}{\kT}$ & $\Ect = E_\mathrm{EL,max}^{~0-0} + \reorg + \frac{\disord^2}{\kT}$ \\
 & $\sigma_\mathrm{PV,EL}=\sqrt{2\reorg\kT + \disord^2}$ & $\sigma_\mathrm{PV,EL}^{0-0}=\sqrt{2\reorg\kT + \disord^2}$ \\
\end{tabular}
\end{ruledtabular}
\end{table}

Kahle \textit{et al.}\ suggested a combination of the \emph{disordered} and \extdMarcus\ to explain $\EQE$, in which each $\Ect'$ in the normal distribution of disordered CT states was extended by vibrational states of equal frequency (equation~(\ref{eq:kahle_EQE})).\cite{kahle2018interpret} If the electro-optical reciprocity relation is applied, we again find a description for $\EL$ involving both the temperature dependent peak redshift inherited from the \disorderedMarcus, and the asymmetry of reduced spectra from the \extdMarcus: 

\begin{eqnarray}
    &&\EQE \propto \frac{E^{-1}\sqrt{\pi^{-1}}}{\sqrt{4\reorg\kT+2\disord^2}} \nonumber\\
    && \times \sum_{j=0}^\infty \FC ~ \exp\left(- \frac{(\Ect+\reorg+j\hw-E)^2}{4\reorg\kT +2\disord^2} \right) \label{eq:kahle_EQE} \\ 
    &&\EL \propto \frac{E \exp{\left(-\frac{\Ect}{\kT}+\frac{\disord^2}{2(\kT)^2}\right)}}{\sqrt{\pi}\sqrt{4\reorg\kT+2\disord^2}} \sum_{j=0}^\infty \FC  e^{\textstyle\left(-j\frac{\hw}{\kT}\right)} \nonumber\\
    &&\times  ~ \exp\left(-\frac{(\Ect +j\hw - \reorg - \disord^2/\kT-E)^2}{4\reorg\kT + 2\disord^2}\right) . \label{eq:kahle_EL}
\end{eqnarray}

This \emph{extended disordered Marcus model} was discussed to be a simplification of a system with two characteristic vibrations; the second vibration was assumed to be in the range of $\hwLow \approx 5\ldots 20\unit{meV}$.\cite{kahle2018interpret, jortner1976temperature} The hypothetical vibration would, in connection with a high Huang--Rhys factor $S_2$, transpose the Poisson distribution at room temperature into a Gaussian shape that is indistinguishable from a normally distributed, statically disordered $\Ect$ at room temperature.\cite{tvingstedt2020temperature} At lower temperatures, this approximation becomes inaccurate, as the mode can no longer be treated in a classical limit, but rather as an additional discrete quantum mode.\cite{ulstrup1975effect} Respective expressions for $\EQE$\cite{kahle2018interpret} and reciprocity compliant $\EL$ of this \multiVibrations are given by:

\begin{widetext}
\begin{eqnarray}
    \EQE \propto &&\frac{E^{-1}}{\sqrt{4\pi\reorg\kT}} ~~~~\sum_{i=0}^\infty \sum_{j=0}^\infty \frac{e^{-S_2} S_2^i}{i!} \FC \exp\left(-\frac{(\Ect + \reorg + j\hw + i\hwLow - E)^2}{4\reorg\kT}\right) \label{eq:twoVibr_EQE} \\ 
    \EL \propto &&\frac{E~e^{\textstyle\left(-\Ect/\kT\right)}}{\sqrt{4\pi\reorg\kT}}  \sum_{i=0}^{\infty} \sum_{j=0}^\infty \frac{e^{-S_2} S_2^i}{i!} \FC e^{\textstyle\left(-\frac{i\hwLow+j\hw}{\kT}\right)} ~ \exp\left(-\frac{(\Ect - \reorg + j\hw + i\hwLow - E)^2}{4\reorg\kT} \right) .\label{eq:twoVibr_EL}
\end{eqnarray}
\end{widetext}

The idea of allowing only dynamic disorder effects for CT lineshape broadening may offer a solution for the $\EL$ peak shift expected for statically disordered systems, but experimentally not observed in OSCs. Yet interestingly, all models that deal with a progression of vibrations also predict an $\EL$ peak redshift with decreasing temperature (albeit not as pronounced as with static disorder). Injected charge carriers will occupy states near the CT state minimum, which is represented in equation~(\ref{eq:twoVibr_EL}) by an exponential decrease of $\EL$ intensity with $i$ and $j$. For a high frequency vibration $\hw \gg \kT$, this results in a vanishing emission from $j>0$; for a low frequency vibration $\hwLow \approx \kT$, it manifests in a perceived peak shift and a further linewidth reduction. As a result of both properties, emission and absorption linewidths will be inherently asymmetrical in this model. A comparison of predicted peak properties between the \emph{multiple vibrations} and \emph{extended disordered model} is shown in Fig.~S4 in the Supplemental material.

\subsection{Consequences}

The main parameters $\Ect$ and $\reorg$ for every model are summarized in table~\ref{tab:table1}, with regard to the linewidths $\sigma$ and peak positions $E_\mathrm{max}$ of reduced EL and PV spectra, wherever possible. While all models share more or less the same set of molecular parameters to describe emission and absorption of CT states, their quantification from experimental data may differ depending on the model of choice. This holds especially true for the dynamically disordered \multiVibrations, where the more complex description prevents a simple extraction of $\Ect$ and $\reorg$. Therefore, even a set of temperature dependent $\EQE$ and $\EL$ measurements requires comprehensive model fits to the spectral data to determine these parameters correctly.

\section{Results}

\subsection{Methods}

We investigate a set of organic bulk heterojunction solar cells with the small molecule donors 1,1-bis[4-(N,N-di-ptolylamino) phenyl]cyclohexane (TAPC) and  4,4',4"-Tris(carbazol-9-yl)triphenylamine (TCTA), diluted with 5 or 10 weight percent (wt\%) in a \fulleren\ matrix. From $\EQE$ measurements under different bias voltages, it can be concluded that these devices have a constant $\IQE$, which is a prerequisite for the electro-optical reciprocity relation.\cite{collado2019energy} The electron mobility in this kind of OSC is several orders of magnitude higher than the hole mobility.\cite{melianas2017charge} In TAPC$\mathrm{_{6\%}}$:\fulleren, the hole mobility is reportedly half of TAPC$\mathrm{_{10\%}}$:\fulleren, yet almost a factor 10 higher than in TCTA$\mathrm{_{6\%}}$:\fulleren, and is further reduced at lower temperatures.\cite{spoltore2018hole} 
Temperature dependent measurements are realized with the devices in a contact gas cryostat. $\EQE$ is measured by comparing currents under short-circuit conditions from monochromatic illuminated OSCs with a calibrated reference detector. Relative $\EL$ is detected from constant current injection, and illumination dependent $\Voc$ is recorded while varying the output power of a cw--laser. Details about the device architecture and experimental conditions are given in sec.~\ref{sec:experimentalSection}.

By choosing a model system of wide bandgap, small molecule donors diluted in a \fulleren\ acceptor matrix, we were able to isolate the absorption of CT states in $\EQE$ spectra, and avoided superimposing donor emission during EL measurements. 

\subsection{Reciprocity and temperature validation}

\begin{figure*}[tb]
    \centering
    \includegraphics[width=\textwidth]{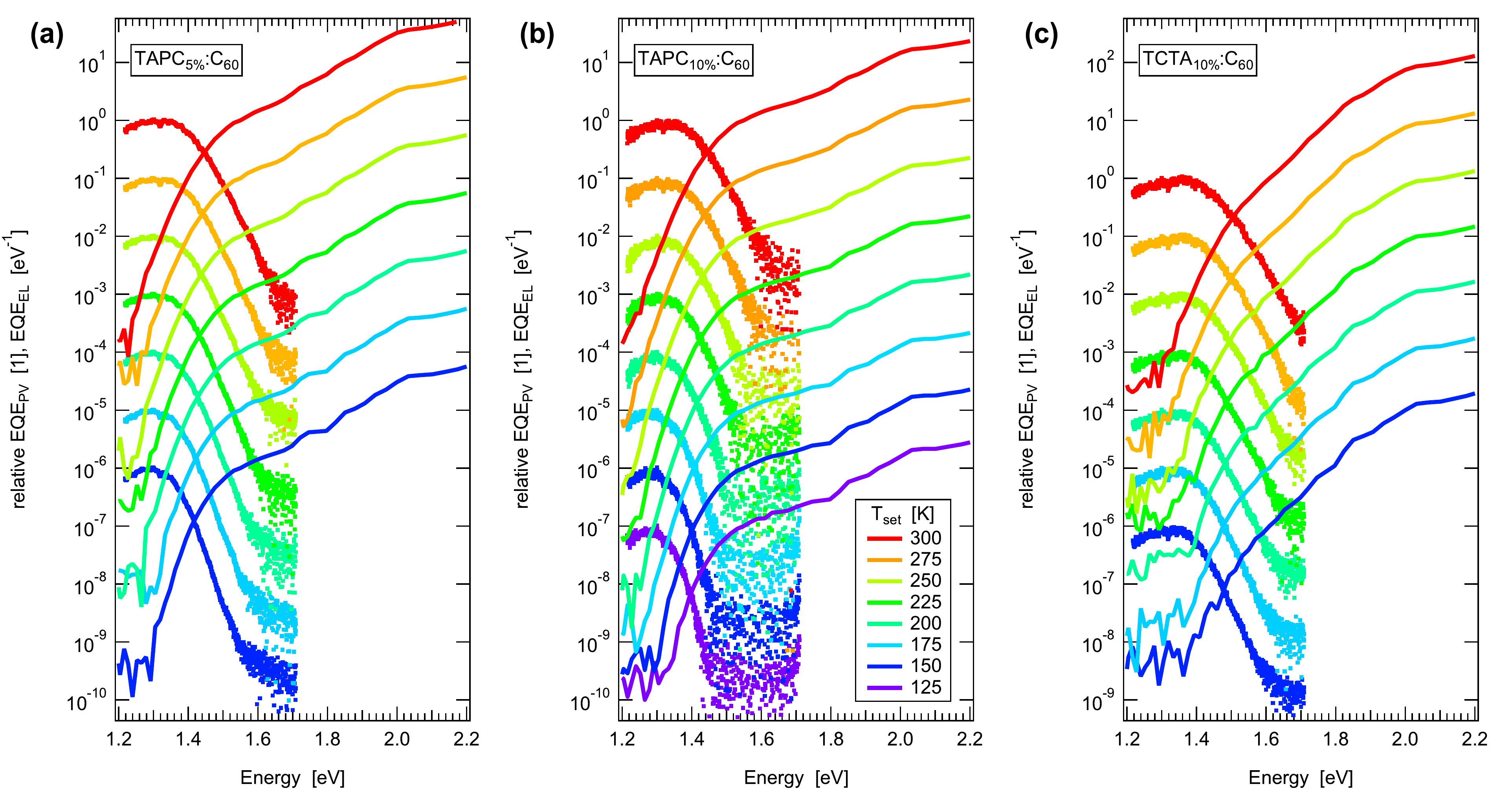}
    \caption{Measured relative $\EQE$ and $\EL$ spectra for (a) TAPC:\fulleren\ bulk heterojunction solar cells with 5 wt\% and (b) 10 wt\% donor content, and (c) TCTA:\fulleren\ bulk heterojunction solar cell with 10 wt\% donor content. All measured curves were shifted by a constant offset respective to $\Tset$. An alternate view on the normalized spectra is given in Fig.~S9 in the Supplemental material.
    }
    \label{fig:input_data}
\end{figure*}

The recorded, temperature dependent $\EQE$ and $\EL$ spectra are shown in Fig.~\ref{fig:input_data}. While the shapes of $\EQE$ spectra are not significantly different at first glance, apart from the decreased signal-to-noise ratio at low temperatures, the deviations are more pronounced in the $\EL$ spectra of each system, most notably by a significant reduction of the linewidth with decreasing temperature.

\begin{figure}[tb]
    \centering
    \includegraphics[width=0.5\textwidth]{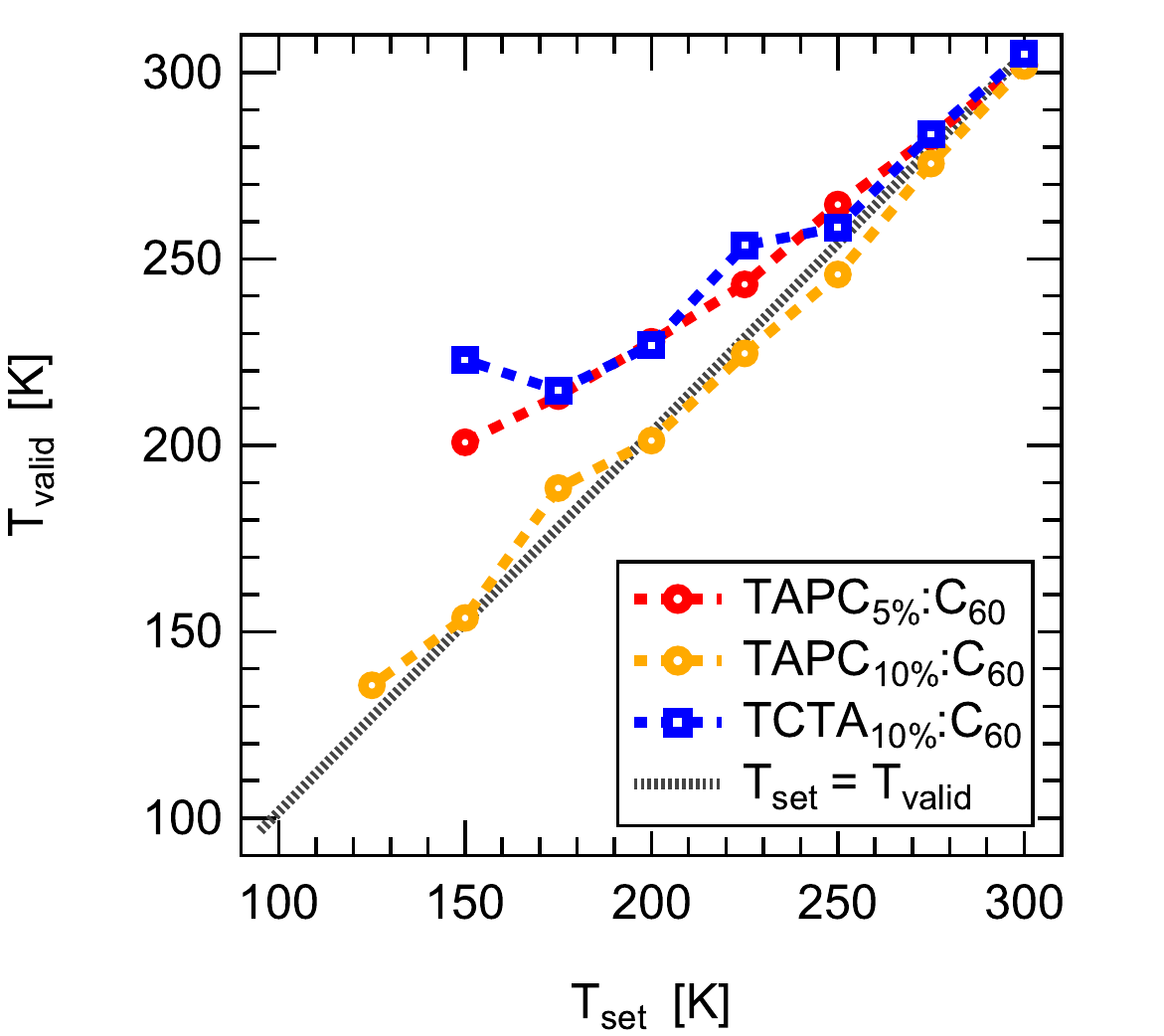}
    \caption{Device temperature $\Tvalid$ obtained from the electro-optical reciprocity relation between $\EQE$ and $\EL$ (equation~(\ref{eq:inverseReciprocity})) with respect to the experimentally expected temperature $\Tset$. For the cell with 10\% donor content, both temperatures are almost identical, whereas we see significant deviations from the ideal case for the other two cells at lower $\Tset$. }
    \label{fig:Tdev_vs_Tmeas}
\end{figure}

First, we checked the validity of the electro-optical reciprocity by extracting $\Tvalid$ from $\EQE$ and $\EL$ spectra measured at the same $\Tset$. The results are shown in Fig.~\ref{fig:Tdev_vs_Tmeas}. We find a very good agreement between $125\unit{K}$ and $300\unit{K}$ for the TAPC:\fulleren\ solar cell with 10\% donor content; however, an increasingly higher $\Tvalid$ is extracted for the other devices when cooled below $250\unit{K}$. 

To further investigate this discrepancy, we compared $\EL$ spectra at presumably equal $\Tset$ (Fig.~S1) and the effect of the injection current density on $\Tvalid$ (Fig.~S2) for the TAPC:\fulleren\ devices. Both control analyses show an increased device temperature during $\EL$ measurements, as the increased $\Tvalid$ correlates with broader $\EL$ spectra, and the effect becoming more pronounced at higher injection currents. We point out that we did not change the experimental routines or the setup between individual devices. Thus, a control mechanism for the device temperature as given by the reciprocity becomes even more important. 

\subsection{Single spectrum analysis}

\begin{figure}[tb]%*
	\includegraphics[width=.5\textwidth]{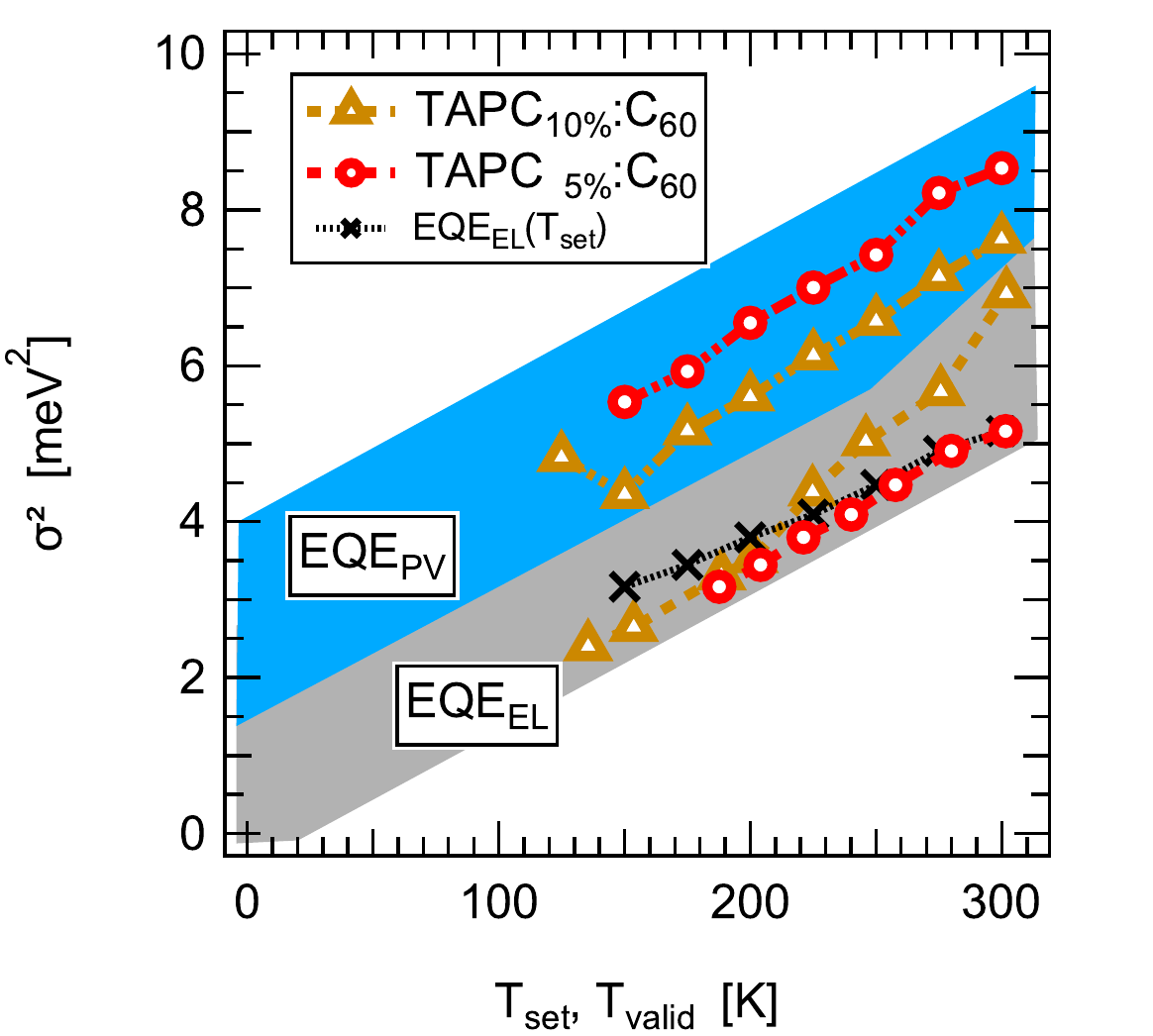}
    \caption{Temperature dependency of the squared CT linewidth $\sigma^2$ in $\EL$ (against validated EL temperature $\Tvalid$) and $\EQE$ (against expected device temperature $\Tset$) spectra for TAPC:\fulleren\ solar cells. We find smaller emission linewidths for all devices. In the case of TAPC$\mathrm{_{5\%}}$:\fulleren, where we found $\Tvalid$ significantly increased for lower $\Tset$, the linewidth narrowing appears less steep if $\Tset$ would be used as reference (black crosses). } 
    \label{fig:linewidths}
\end{figure}

When analyzing the linewidth $\sigma$ of $\EL$ and $\EQE$ CT contributions individually in terms of a Gaussian distribution, currently the most common method,\cite{linderl2020crystalline, tvingstedt2020temperature} we find major differences between extracted values from $\EQE$ and $\EL$ spectra (Fig.~\ref{fig:linewidths}; values for TCTA$_\mathrm{10\%}$:\fulleren\ in Fig.~S5). Rather than fitting a non-linear function to the spectrum, we applied a linear regression to the numerical derivative $\mathrm{d/d}E$ of the reduced spectra to decrease fitting errors\cite{tvingstedt2020temperature} (see sec.~5 in the Supplemental material). In general, our $\EQE$ spectra tend to be significantly broader than their $\EL$ counterparts, requiring an asymmetrical model to describe both spectra simultaneously. These findings show that CT state models requiring reduced emission and absorption spectra to be symmetrical---the \textit{simple} and \textit{disordered Marcus model}---cannot be correct.\cite{vandewal2010relating, burke2015beyond} 
 
%*

The slope of $\sigma^2(T)$ and its extrapolated intercept at $T=0\unit{K}$ is used to extract values for $\disord$ and $\reorg$ based on the \emph{disordered} and \emph{extended disordered Marcus model} (equations~(\ref{eq:burke_EQE})-(\ref{eq:kahle_EL})).\cite{tvingstedt2020temperature,burke2015beyond, linderl2020crystalline} The extracted parameters are listed in table~\ref{tab:DisorderReorganization}. Both parameters depend on whether they are based on the supposed ($\Tset$) or validated temperature ($\Tvalid$) during the measurement. Negative values for $\disord$ indicate a negative interpolated intersection which would be unphysical in terms of the \disorderedMarcus. Similar results were also reported for low-temperature EL measurements, challenging the relevance of static energetic disorder for the CT state ensemble compared to dynamic broadening.\cite{tvingstedt2020temperature}  
In addition, we find unexpected quantitative differences of the parameters between absorption and emission, and within similar donor dilutions. Instead of relying on individual $\EQE$ or $\EL$ peak analysis with all its flaws, our temperature validated combined data records allow us to extract model parameters from a simultaneous review of absorption and emission spectra. 

\begin{table}
	\footnotesize
	\caption{\label{tab:DisorderReorganization}Energetic width of static disorder distribution $\disord$ and reorganization energy $\reorg$ extracted from individual fits to reduced $\EQE$ and $\EL$ spectra in accordance with the \disorderedMarcus. Negative values for $\disord$, noted in parenthesis, occur when the CT linewidth interpolates to negative values at $T=0\unit{K}$. Both obtained parameters change significantly when the correction of $\Tset$ is applied for $\EL$ by utilizing the reciprocity relation.} 
	\begin{ruledtabular}
	\setlength{\tabcolsep}{0em}
		\begin{tabular}{cc|cc|cc|}
		 & & \multicolumn{2}{c|}{$\EL$} & \multicolumn{2}{c|}{$\EQE$} \\
		 & & $\disord$/meV & $\reorg$/meV & $\disord$/meV & $\reorg$/meV \\ \hline
		 \multirow{2}{*}{TAPC$_\mathrm{5\%}$:\fulleren} & $\Tvalid$ & (-14.8) & 104.8 & \multirow{2}{*}{49.7} & \multirow{2}{*}{117.7} \\ 
		  & $\Tset$ & 32.6 & 79.6 &  &  \\ \hline
		 \multirow{2}{*}{TAPC$_\mathrm{10\%}$:\fulleren} & $\Tvalid$ & (-38.9) & 151.4 & \multirow{2}{*}{36.7} & \multirow{2}{*}{122.2}  \\
		  & $\Tset$ & (-33.3) & 145.9 &  &  \\ \hline
		 \multirow{2}{*}{TCTA$_\mathrm{10\%}$:\fulleren} & $\Tvalid$ & (-41.3) & 179.3 & \multirow{2}{*}{44.9} & \multirow{2}{*}{126.7}  \\ 
		  & $\Tset$ & 26.0 & 134.5 & & \\
		\end{tabular}
	\end{ruledtabular}
\end{table}

\subsection{Global data analysis}

To test the \multiVibrations, we implemented a Levenberg-Marquardt optimizing algorithm to fit both $\EQE$ and $\EL$ spectra simultaneously for multiple temperatures. As discussed before, we assumed the temperature $\Tvalid$ validated by the reciprocity relation for $\EL$-measurements, while maintaining $\Tset$ for $\EQE$-spectra. Each spectrum was weighted equally by dividing its residuals by the number of supporting data points, in order to avoid over-emphasizing of $\EL$ (with higher energy resolution than $\EQE$), and of spectra from higher temperatures, which tend to provide more supporting data points due to their better signal-to-noise ration and broader signal range. The fitting range of models that do not describe excitation of molecular vibrations was reduced to the tail of each spectrum; otherwise, it is limited by the onset of the \fulleren-singlet excitation at $1.7\unit{eV}$ for $\EQE$ and the detection limit at $1.2\unit{eV}$ for $\EL$, and respective background noise levels. We applied the algorithm for all models summarized in table~\ref{tab:table1} with similar starting values. During the fit, the parameter describing the high frequency vibration was held constant at the lower limit ($\hw=150\unit{meV}$) of the values reported for the carbon-carbon stretching mode.

\begin{figure*}[tb]
    \includegraphics[width=\textwidth]{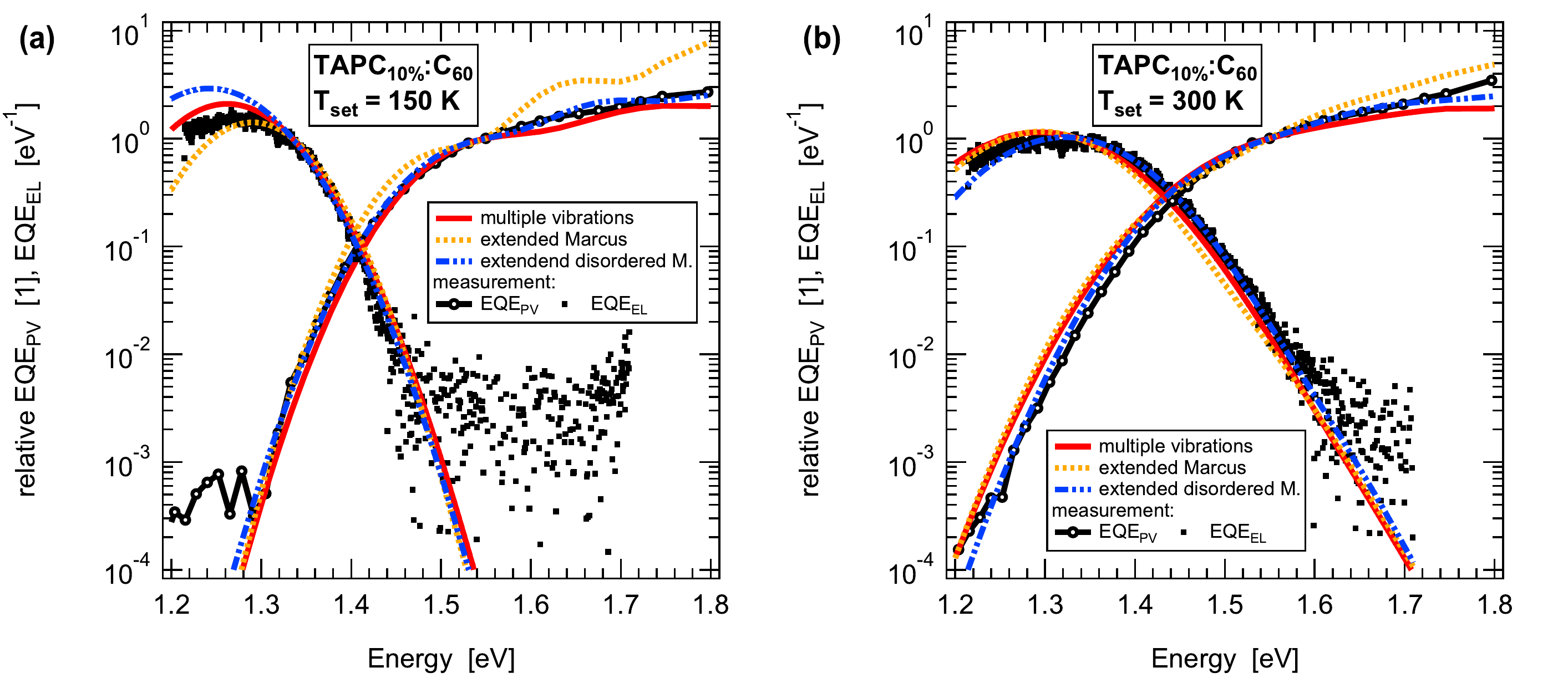} 
    \caption{Exemplary model fits for (a) $T=150\unit{K}$ and (b) $300\unit{K}$ for the solar cell with 10 wt\% TAPC content. We find the best combined agreement for $\EQE$ and $\EL$ at all temperatures with the \emph{multiple vibrations model}. The \emph{extended Marcus model} would require a defined substructure of additional absorption lines, separated by the characteristic vibrations energy $\hw$. Respective peaks are not found at low temperatures, which could be explained by the \emph{extended disordered Marcus model}; however, it would require the $\EL$ peak shifting at low temperatures, which is not observed experimentally. } 
    \label{fig:fit_result}
\end{figure*}

Fig.~\ref{fig:fit_result} shows measured reduced $\EQE$ and $\EL$ spectra of \tapc\ at $T=150\unit{K}$ and $300\unit{K}$ and respective fitted curves for the \emph{multiple vibrations model, extended} and \emph{extended disordered Marcus model}. The best agreement between data and model is found for the \multiVibrations, while the others may properly fit the spectra at room temperature and their respective tail regions, yet lead to characteristic differences especially at lower temperatures. A more detailed discussion is given in section~IV.

Some values of the fitted CT energy $\Ect$, reorganization energy $\reorg$ and the low-energy vibration ($\hwLow$ and $S_2$) are listed in table~\ref{tab:FitResults}, with the complete sets of parameters in table~S3. We focus here on the more elaborate \emph{multiple vibration}, \emph{extended} and \emph{extended disordered Marcus model}. The latter two models imply a more than $100\unit{meV}$ higher CT state energy $\Ect$ and a lower reorganization energy $\reorg$ of the CT state than the former. Between the different donor concentrations in TAPC:\fulleren\ blends, we find a low energy vibration of around $13\unit{meV}$ and a variation in the related Poisson distribution moment $S_2$.

\begin{table}
	\footnotesize
	\caption{\label{tab:FitResults} Extrapolated open circuit voltage limit $\Voc^{0\mathrm{K}}$ and fitted parameters for the multiple vibration (mv), extended (eM) and extended disordered Marcus model (edM). We systematically find $\Ect$ to be at least $100\unit{meV}$ higher in the extended and extended disordered Marcus model, in correspondence with a significant lower reorganization energy. } 
	\begin{ruledtabular}
		\begin{tabular}{c||c|c|c||c|c|c||c|c|}
		 & \multicolumn{3}{c||}{$\Ect/\mathrm{eV}$} & \multicolumn{3}{c||}{$\reorg/\mathrm{meV}$} & $\hwLow /\mathrm{meV}$ & $S_2$\\
		 & eM&edM&mv & eM&edM&mv & \multicolumn{2}{c|}{mv} \\ \hline
		 TAPC$_\mathrm{5\%}$:\fulleren & 1.52&1.48&1.38 & 67&68&93 & 13.7&2.3 \\ 
		 & \multicolumn{3}{c||}{$q\Voc^{0\unit{K}} = 1.38\unit{eV}$} &&&&& \\
		 \hline
		 TAPC$_\mathrm{10\%}$:\fulleren & 1.47&1.47&1.33 & 71&73&114 & 12.5&8.9 \\
		 & \multicolumn{3}{c||}{$q\Voc^{0\unit{K}} = 1.32\unit{eV}$} &&&&& \\ 
		 \hline
		 TCTA$_\mathrm{10\%}$:\fulleren & 1.45&1.60&1.34 & 51&62&132 & 7.1&18.2 \\
		 & \multicolumn{3}{c||}{$q\Voc^{0\unit{K}} = 1.47\unit{eV}$} &&&&& \\
		\end{tabular}
	\end{ruledtabular}
\end{table}

To add further plausibility to the numerically obtained model parameters, we turned to the total EL quantum yield $\LED$, which can be estimated by integration of measured and predicted $\EL$ spectra. We find good agreement for the models without static energetic disorder, while the calculated $\LED$ in the \extdDisorderedMarcus\ becomes significantly larger than measured values at low temperatures (Fig.~S8). 
In addition, we measured $\Voc$ for a range of temperatures and increasing illumination intensity $G$; values and analysis are discussed in sec.~6 in the Supplemental material. Regardless of the intensity, $\Voc(T)$ converges to a fixed limit when extrapolated to $T\rightarrow0\unit{K}$. The limit $q\Voc^{0\unit{K}}$, shown in table~\ref{tab:FitResults}, should equal $\Ect$ in the case of OSCs,\cite{vandewal2010relating} thus providing us with a supplementary measurement to validate the model parameters. 

\section{Discussion}

The agreement between $\Tset$ and $\Tvalid$ for \tapc\ over a temperature range of $175\unit{K}$ implies the validity of the electro-optical reciprocity relation between $\EQE$ and $\EL$, as $\Tvalid$ was obtained by strictly comparing measured spectra taken at $\Tset$. However, we find this correlation to be violated for the other solar cells investigated in this study. If the reciprocity relation remained valid as expected, we would conclude that $\Tset$, the temperature of the heat reservoir, differs from the actual device temperature either during the $\EQE$ or $\EL$ measurement, or both. On the one hand, a lower solar cell temperature while recording $\EQE$ is unlikely as we always decreased $\Tset$ during the experiment and no additional cooling mechanisms were present. On the other hand, we always find $\Tvalid > \Tset$, which leads us to conclude that the solar cell temperature is increased during the $\EL$ measurements: $\Tvalid$ increases with higher injection currents during the EL experiment. In this case, perfect agreement between $\Tvalid$ and $\Tset$ is found only for injection current densities $J_\mathrm{inj} < 10\unit{mA\,cm^{-2}}$ (Fig.~S2). Due to the design of the experiment, in which we recorded consecutive $\EL$ spectra for increasing $J_\mathrm{inj}$ at each temperature $\Tset$ before cooling down to the next temperature step, we can safely assume the solar cells have reached thermal equilibrium with the reservoir before we injected charge carriers. The most likely explanation for this behavior is additional heating of the whole devices induced by electric transport and shunt currents, as suspected previously.\cite{vandewal2010relating}

The reported discrepancy between hole and electron mobilities in diluted donor:\fulleren OSCs might lead to accumulation of space charges in the devices at lower temperatures, which would violate a basic assumption of the reciprocity relation. This effect should be less pronounced in TAPC$\mathrm{_{10\%}}$:\fulleren due to its higher hole mobility at room temperature compared to the other devices.\cite{spoltore2018hole} However, it was shown that inbalanced mobilities have a negligible influence on reciprocity in thin OSCs.\cite{kirchartz2016reciprocity} Due to the active layer thickness of only $50\unit{nm}$, and the apparently unaffected reciprocity at $300\unit{K}$ and under lower injection conditions in 5\% TAPC, space charges seem unlikely as the main cause of the temperature difference. At the same time, a lower mobility implies lower conductivity and increased Joule heating.

We applied the common method of fitting reduced Gaussian distributions to sub-bandgap $\EQE$ or $\EL$ tails to determine energetic CT state parameters $\Ect$, $\disord$ and $\reorg$ according to the \emph{disordered Marcus model} (Tab.~\ref{tab:DisorderReorganization}). Even though we used a linear regression method instead of non-linear fitting to reduce the uncertainties related to manual adjustment of fit ranges and starting values,\cite{tvingstedt2020temperature} the extracted parameter values from individual $\EQE$ and $\EL$ analyses were inconsistent, or outright unphysical within the model. 

The systematic discrepancy between apparent CT state linewidths in emission and absorption leads us to the conclusion that reduced $\EQE$ and $\EL$ distributions \emph{are} in fact asymmetrical. This rules out the \textit{simple} and \textit{disordered Marcus model}, both predicting symmetry, to describe emission and absorption by CT states. Instead, the experimental observations are compatible with the \textit{multiple vibrations} model. As was described before,\cite{tvingstedt2020temperature} the Poisson distributions attributed with low energy vibrations merge into reduced Gaussians at high temperatures. Yet, if incorrectly interpreted as Gaussians, the extracted parameters could be misleading, and may even implicate apparent negative widths $\disord$ of the assumed static disordered $\Ect$ distributions. 

To analyze our measured data with the \multiVibrations, we reconstructed the measured $\EQE$ and $\EL$ spectra for all temperatures simultaneously by a Levenberg--Marquard fitting algorithm (Fig.~\ref{fig:fit_result}). A similar fit featuring the \emph{extended} and \emph{extended disordered Marcus model} resulted in further arguments that these two models do describe the physics of the investigated devices. The distinct substructure of higher energetic molecular vibration peaks at low temperatures could not be replicated in the measured, comparably smooth $\EQE$ spectra. To compensate the missing substructure, static energetic disorder would be required; however, the static \emph{disorder models} postulate a significant $\EL$ peak shift below the low-energy edge of the fitting range at lower temperatures. This leads to increasing peak amplitudes as well as fit residuals in this region, and the estimated $\LED$-values correspondingly deviate from the integrated measured $\EL$ spectra at lower temperatures. While we cannot provide measured emission data in this region to back up the prediction, we suspect this fitting behavior to solely be a numerical compensation for a temperature dependent peak shift required by the model, which is not present in the experimental data. Either way, both models fail to reconstruct measured spectral data as a whole, which raises further doubt about their validity. 

The \emph{multiple vibrations model} yields a much more reasonable reconstruction of the experimental spectra. We even find agreement between the values of measured and reconstructed integrated $\LED$ (Fig.~S8), and within the extracted parameter quantities for the donor material TAPC in different dilutions (table~\ref{tab:FitResults}).

The choice of the model to analyze measured data obviously has an effect on the extracted parameters values, and we do not have to look further than the excited CT state energy $\Ect$ to see the importance of the model choice. In most discussed models, $\Ect$ is determined by the $\EQE$'s (0--0) peak position and the reorganization energy $\reorg$ (values in table~S3). This corresponds to the intersection between reduced normalized emission and absorption peak\cite{vandewal2010relating} if static energetic disorder is not present; else, we would see a temperature dependent emission peak shift as discussed earlier. 

\begin{figure}[tb]
	\includegraphics[width=.49\textwidth]{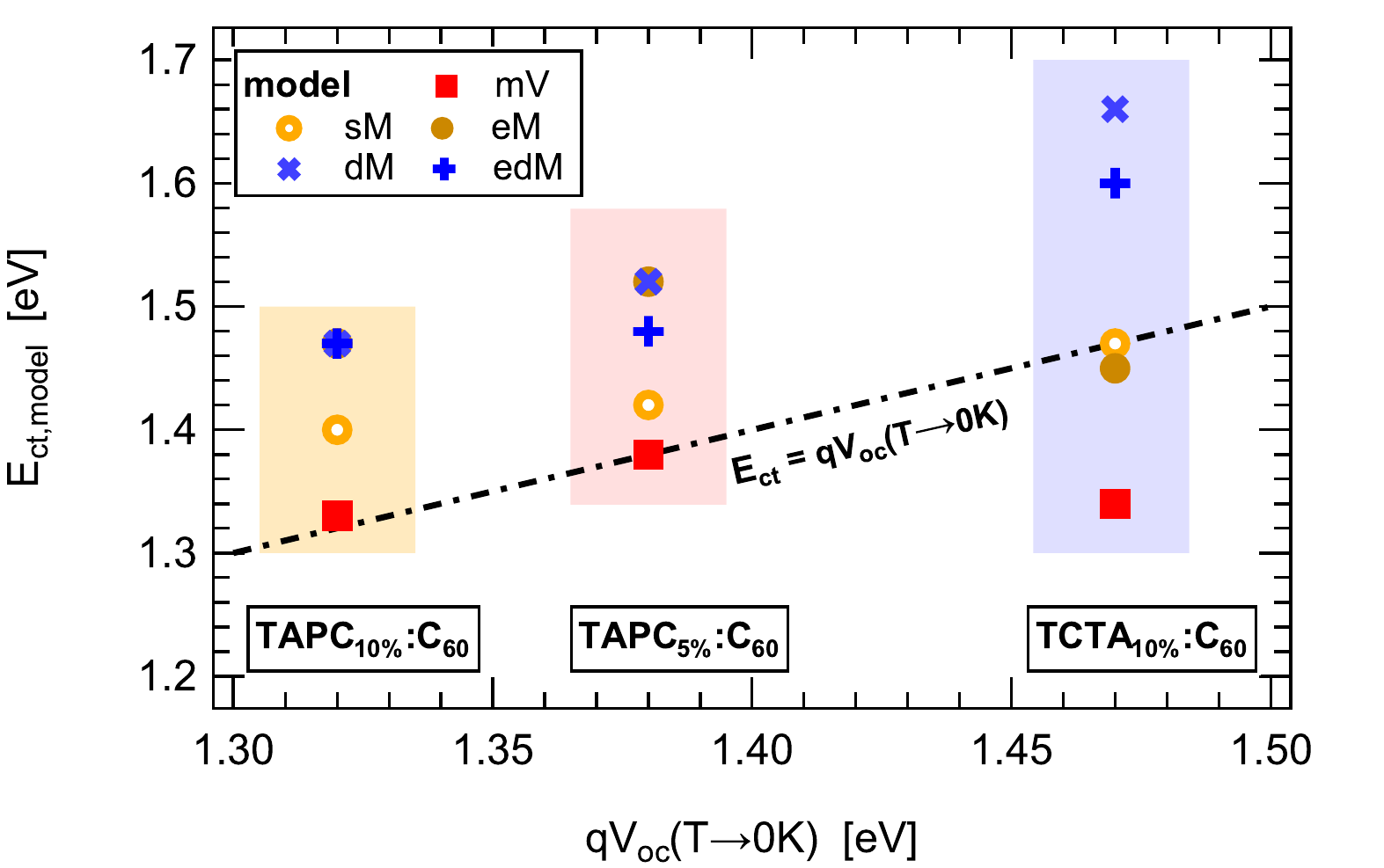}
    \caption{Fitted energies $E_\mathrm{ct,model}$ of the CT state according to the simple (sM), extended (eM), disordered (dM), and extended disordered Marcus model (edM), and the multiple vibrations model (mV), compared to the extrapolated open circuit voltage limit $\Voc(T\rightarrow0\unit{K})$. The black line represents identity, which is best approximated by the multiple vibrations model for TAPC:\fulleren\ blends. Models which include static energetic disorder (dM, edM) overestimate $\Ect$.} 
    \label{fig:Voc_vs_Ect}
\end{figure}

In the \multiVibrations, the absorption peak center is determined by the moment of the Poisson distribution of vibrational states with frequency $\hwLow$, and thus exceeds $\Ect+\reorg$. Accordingly, the reconstructed values for $\Ect$ are here lower than in the other investigated models. When compared to $\Voc$ extrapolated to $0\unit{K}$, we find them best approximated with the \multiVibrations's lower $\Ect$ values in the case of TAPC:\fulleren\ blends (see Fig.~\ref{fig:Voc_vs_Ect}). In contrast, the static disorder models (\textit{disordered} and \textit{extended disordered Marcus model}) overestimate the limit by more than $100\unit{meV}$ (table~\ref{tab:FitResults}).

Here, we want to highlight that the most trusted results of our analysis are gathered from the TAPC$_\mathrm{10\%}$:\fulleren\ device, as this is the one that has been measured at the correct emission temperature, i.e., at $\Tvalid = \Tset$. In all other devices, we were able to at least estimate $\EL$ temperatures from the reciprocity relation, which we are convinced of is much closer to the real device temperature than the reservoir temperature $\Tset$, for the reasons discussed earlier. Differences in the extracted parameters, especially between the diluted TAPC solar cells, which only vary in their donor concentration and therefore should not differ a lot in their fundamental energetic properties, may partly be attributed to this temperature uncertainty. From the $\EQE$ and $\EL$ spectra alone, we find that $\Ect$, $\reorg$ and $S_2$ are strongly correlated, in a way that the $\Ect$ difference between the cells with 5 wt\% and 10 wt\% TAPC concentration is compensated by the difference in $\reorg$ and $S_2$, while $\hwLow$ remains almost constant. To verify the parameters even further, additional experimental methods to access $\hwLow$ and $\Ect$ would be necessary.

\section{Conclusion}

We measured the $\EQE$ and $\EL$ properties of diluted donor--acceptor solar cell model systems and tested currently discussed theoretical models concerning the temperature behavior of CT state emission and absorption. The OSCs nonetheless show substantial linewidth contributions independent of temperature (table~\ref{tab:DisorderReorganization}), which are comparable to reported values from state-of-the-art solar cells extracted by either temperature dependent absorption or emission spectroscopy.\cite{kahle2018interpret, tvingstedt2020temperature, linderl2020crystalline, burke2015beyond}

We applied the electro-optical reciprocity relation between $\EQE$ and $\EL$ spectra as a method to validate and correct the emission temperature of the devices. By showing that validated temperatures generally comply with expected temperatures, we confirm the validity of the reciprocity relation for this set of devices. The validated temperature can be several $10\unit{K}$ higher than assumed from the set temperature under different experimental circumstances, such as higher injection current densities during EL measurements. We attributed this discrepancy to additional current-induced heating of the devices, and recommend to carefully verify the emission temperature during EL experiments.

We are missing experimental evidence for a $\EL$ peak shift and find no symmetry between absorption and emission spectra---both predicted by gaussian \textit{static disorder models}. Together with their inability to reconstruct measured data and estimated values for $\LED$, we have reasonable doubt about the general applicability of these models to describe CT state absorption and emission. Non-gaussian distributions of static disorder might be considered instead if they predict both details and the general spectral shape correctly.

Our measured spectra are indeed better described by a \emph{multiple vibrations model} with two characteristic frequencies of around $150\unit{meV}$ and $13\unit{meV}$ for TAPC:\fulleren, and $7\unit{meV}$ for TCTA:\fulleren. Higher energy modes above $30\unit{meV}$ have been reported for \fulleren.\cite{tvingstedt2020temperature} Because we found similar frequencies for TAPC, and a different one for TCTA, it is a likely scenario that the low frequency vibration is associated with the donor. As a consequence, we find the excited CT state energy to be significantly smaller than suggested by previous analyses of TAPC and TCTA:\fulleren\ blends.\cite{benduhn2017intrinsic} We validated the \emph{multiple vibrations model} further by showing agreement between extrapolated open circuit voltage limit at $T=0\unit{K}$ and extracted values for $\Ect$ for TAPC:\fulleren\ devices. 

In conclusion, while static energetic disorder in combination with the \multiVibrations might play a role in different donor:acceptor blends, dynamic disorder has to be addressed in all organic solar cell systems as a dominant mechanism.

\section{Experimental section}
\label{sec:experimentalSection}

\emph{Materials:} 
1,1-bis[4-(N,N-di-ptolylamino) phenyl]\-cyclohexane (TAPC) and 4,4’,4"-Tris(carbazol-9-yl)triphenylamine (TCTA) were purchased from Sensient Technologies Corporation (USA), \fulleren\ from CreaPhys GmbH (Germany), and Luminescence Technology Corp. (Lumtec, Taiwan); 4,7-diphenyl-1,10-phenanthroline (BPhen) was purchased from Abcr GmbH (Germany) and Luminescence Technology Corp., Molybdenum trioxide from Luminescence Technology Corp.

\emph{Device fabrication:} The devices studied in this work are fabricated layer by layer through thermal co-evaporation in ultra-high vacuum chamber (K. J. Lesker, UK) which has typical operating pressure of $10^{-7}\unit{mbar}$. All of the involved organic molecules are thermally sublimated before evaporation. Substrates are glass substrates (size $25\times25\unit{mm^2}$) with pre-patterned ITO (indium tin oxide, Thin film devices, US). The ITO is $90\unit{nm}$ thick with a sheet resistance of $25\unit{\Omega cm^{-2}}$, with 84\% transparency. ITO is covered with $2\unit{nm}$-thick hole transport layer of Molybdenum trioxide. The co-evaporated donor--acceptor blends of TAPC:\fulleren\ and TCTA:\fulleren\ with weight ratios of 5:95 and 10:90 is further covered with $8\unit{nm}$-thick layer of the electron transport material BPhen and a $100\unit{nm}$ aluminum top contact. The active area of the device is defined by the intersection of the structured ITO electrode and the structured opaque aluminum top contact and amounts to $6.44\unit{mm^2}$. Before device fabrication, substrates are cleaned with following procedure: coarse cleaning by detergent; rinsing with de-ionized (DI) water; sequentially dipping into N-Methyl-2-pyrrolidone (NMP), acetone, ethanol for ultrasonic bath with $8\unit{min}$. for each solvent; rinsing with de-ionized water; after drying up, oxygen plasma (Priz Optics, Germany) treatment for 10 min. After fabrication in ultra-high vacuum, devices are transferred into a nitrogen filled glovebox. As the final step, all devices are encapsulated with a small glass lid to prevent moisture and oxygen induced degradation during device characterization in air ambient. The transparent glass lid is glued by UV-(ultra violet)-light-curing epoxy resin (XNR 5592, Nagase ChemteX, Japan) which is exposed with UV light for $196\unit{s}$.

\emph{External quantum efficiency}: $\EQE$ is measured inside a closed-cycle cryostat (Cryovac) with helium as the contact gas. The monochromatic light source consists of a mechanically chopped $100\unit{W}$ quartz-tungsten-halogen lamp coupled to a double monochromator with additional optical bandpass filters (Quantum Design Europe MSHD-300) for further stray light reduction. The photocurrent, measured under short circuit conditions and without bias illumination, is amplified and converted to a voltage signal by a trans-impedance amplifier (Zurich Instruments HF2TA) and measured with a lock-in amplifier (Zurich Instruments HF2LI). A small fraction of the monochromatic light is directed onto a calibrated two-color photo-diode (Hamamatsu K1718-B) to record reference excitation fluxes with an additional lock-in amplifier (Stanford Research DSP830). 

\emph{Electroluminescence:} $\EL$ spectra are recorded with a $500\unit{mm}$ spectrograph (Princeton Instruments Acton 500i) and a liquid nitrogen-cooled silicon CCD-camera (Princeton Instruments Spec-10:100). The constant EL driving current of $155\unit{mA\,cm^{-2}}$ is provided by a source--measure unit (Keithley 2368B).

\emph{Open circuit voltage:} A frequency doubled Nd:YAG-LASER (Spectra Physics Millenia Pro) with optical output power of $1\unit{W}$ is used to illuminate OSCs inside the custom-build cryostat. The incident irradiance is gradually increased over 10 orders of magnitude by passing two motorized neutral-density filter wheels (Thorlabs FW102C \& FW212C); $\Voc$ of the illuminated device is measured with a source--measure-unit (Keithley 2368B). A mechanical shutter, triggered by the source--measure unit, limits the illumination time of the devices.

\begin{acknowledgments}
We acknowledge funding by the DFG (project DE830/19-1). The work of M.S.\ received funding from the European Union’s Horizon 2020 research and innovation program under the Marie Skłodowska-Curie Grant Agreement No.\ 722651 (SEPOMO). J.B.\ acknowledges the Sächsische Aufbaubank through project no. 100325708 (Infrakart). We want to thank Lukasz Baisinger (IAPP Dresden) for his support during device fabrication and initial characterization.
\end{acknowledgments}

\bibliography{goehler2021temperature}

\end{document}